\documentclass{PoS}
\title{Decay constants of heavy mesons from QCD sum rules}

\ShortTitle{Decay~constants~of~heavy~mesons~from~QCD~sum~rules}

\author{Wolfgang Lucha\\ HEPHY, Austrian Academy of Sciences,
Nikolsdorfergasse 18, A-1050 Vienna, Austria\\ E-mail:
\email{wolfgang.lucha@oeaw.ac.at}}
\author{\speaker{Dmitri Melikhov}\\ HEPHY, Austrian Academy of
Sciences, Nikolsdorfergasse 18, A-1050 Vienna, Austria\\ Faculty
of Physics, University of Vienna, Boltzmanngasse 5, A-1090 Vienna,
Austria\\ SINP, Moscow State University, 119991, Moscow, Russia\\
E-mail: \email{dmitri\_melikhov@gmx.de}}
\author{Silvano Simula\\ INFN, Sezione di Roma III, Via della Vasca
Navale 84, I-00146, Roma, Italy\\ E-mail:
\email{simula@roma3.infn.it}}

\abstract{We present a sum-rule extraction of the decay constants
of the $D$, $D_s$, $B$, and $B_s$ mesons from the two-point
correlator of heavy-light pseudoscalar currents \cite{lms2010}.
Our main emphasis is laid on the control over the uncertainties in the
decay constants, related both to the input QCD parameters and to
the limited accuracy of the method of sum rules. Gaining this
control has become possible due to the application of our novel 
procedure for extracting hadron observables based on a dual
threshold depending on the Borel parameter. For charmed mesons, we
obtain $f_D=(206.2\pm 7.3_{(\rm OPE)}\pm 5.1_{(\rm syst)})$ MeV and
$f_{D_s}=(245.3\pm 15.7_{(\rm OPE)}\pm 4.5_{(\rm syst)})$ MeV. For
beauty mesons, the decay constants turn out to be extremely
sensitive to the precise value of the $\overline{\rm MS}$ mass of the $b$-quark, 
$\overline{m}_b(\overline{m}_b)$. By requiring our sum-rule
estimate to match the average of lattice determinations of $f_B$,
a very accurate value is extracted,
$\overline{m}_b(\overline{m}_b)=(4.245\pm 0.025)$ GeV, yielding $f_{B} = (193.4 \pm 12.3_{\rm (OPE)} \pm
4.3_{\rm (syst)})\; {\rm MeV}$ and $f_{B_s} = (232.5 \pm 18.6_{\rm
(OPE)} \pm 2.4_{\rm (syst)})\; {\rm MeV}$.}

\FullConference{35th International Conference of High Energy
Physics -- ICHEP2010,\\ July 22--28, 2010\\ Paris France}

\begin{document}
\section{Introduction}
The extraction of the decay constants of ground-state heavy pseudoscalar mesons 
within the method of QCD sum rules \cite{svz} is~a complicated problem:
First, one should derive a reliable operator product expansion
(OPE) for the Borelized correlation function of two
pseudoscalar heavy-light currents. We make use~of~the OPE for this
correlator to three-loop accuracy \cite{chetyrkin}, reshuffled in
terms of the $\overline{\rm MS}$ heavy-quark mass, in which case
the perturbative expansion exhibits a reasonable convergence
\cite{jamin}.
Second, the knowledge of the truncated OPE for the correlator ---
even if the parameters of this OPE are known precisely --- allows
extracting the characteristics of the bound state with a limited accuracy 
which reflects the intrinsic uncertainty of the method of
QCD sum rules. The control over this uncertainty is a
very subtle problem \cite{lms_2ptsr}.

Recently, we have formulated
a new approach to extracting the ground-state parameters from the
correlator which enables such a control \cite{lms_new}.~Let~us
briefly recall the essential features of our approach: The
quark--hadron duality assumption leads~to~a relation between
ground-state contribution and OPE for the correlator with a cut at some effective continuum
threshold $s_{\rm eff}$:
\begin{eqnarray}
\label{SR_QCD} 
f_Q^2 M_Q^4 e^{-M_Q^2\tau}=\Pi_{\rm dual}(\tau,
s_{\rm eff}(\tau)) \equiv \int\limits^{s_{\rm
eff}(\tau)}_{(m_Q+m)^2} ds \, e^{-s\tau}\rho_{\rm pert}(s) +
\Pi_{\rm power}(\tau).
\end{eqnarray}
Evidently, in order to extract the decay constant one has to fix
the effective continuum threshold~$s_{\rm eff}$. Moreover, as is
obvious from (\ref{SR_QCD}) $s_{\rm eff}$ must be a function of
$\tau$, otherwise the l.h.s.\ and the r.h.s.\ of~(\ref{SR_QCD})
exhibit a different $\tau$-behaviour. The {\it exact effective
continuum threshold}, corresponding~to exact values of
hadron mass and decay constant on the l.h.s., is of course not
known.~Therefore, the extraction~of~hadron parameters from the sum rule
consists in attempting (i) to find~a reasonable 
approximation to the exact threshold and (ii) to control the
accuracy of this approximation.

Let us introduce the dual invariant mass $M_{\rm dual}$ and the
dual decay constant $f_{\rm dual}$ by the relations
\begin{eqnarray}
\label{mdual} M_{\rm dual}^2(\tau)&\equiv&-\frac{d}{d\tau}\log
\Pi_{\rm dual}(\tau, s_{\rm eff}(\tau)),
\\
\label{fdual} f_{\rm dual}^2(\tau)&\equiv&M_Q^{-4}
e^{M_Q^2\tau}\Pi_{\rm dual}(\tau, s_{\rm eff}(\tau)).
\end{eqnarray}
If the mass of the ground state is known, any deviation of the
dual mass from the actual mass of the ground state yields an
indication of the excited-state contributions picked up by the
dual correlator. Assuming a particular functional form of the
effective threshold and requiring the least deviation of the dual
mass (\ref{mdual}) from the actual mass in the
$\tau$-window leads to a variational solution for the effective
threshold; as soon as the latter has been fixed we derive the
decay constant~from~(\ref{fdual}). The standard assumption
for the effective threshold is a $\tau$-independent constant. In
addition~to this approximation, we have considered polynomials in
$\tau$. Reproducing the actual mass considerably improves for
$\tau$-dependent thresholds. This means that a dual correlator
with $\tau$-dependent threshold isolates the ground-state
contribution much better and is less contaminated by the excited
states~than a dual correlator with standard $\tau$-independent
threshold. As consequence, the accuracy of extracted hadron
observables improves considerably. Recent experience from potential models reveals that the band of
values obtained from the linear, quadratic, and cubic Ans\"atze
for~the effective threshold encompasses the true value of the
decay constant \cite{lms_new}. Moreover, we could show that the
extraction procedures in quantum mechanics and in QCD are even
quantitatively rather similar~\cite{lms_qcdvsqm}. 

This contribution reports our recent results \cite{lms2010} for the heavy-meson decay constants.
\section{Decay constants of the $D$ and $D_s$ mesons}
\begin{figure}[!ht]
\begin{center}
\begin{tabular}{cc}
\includegraphics[height=7cm]{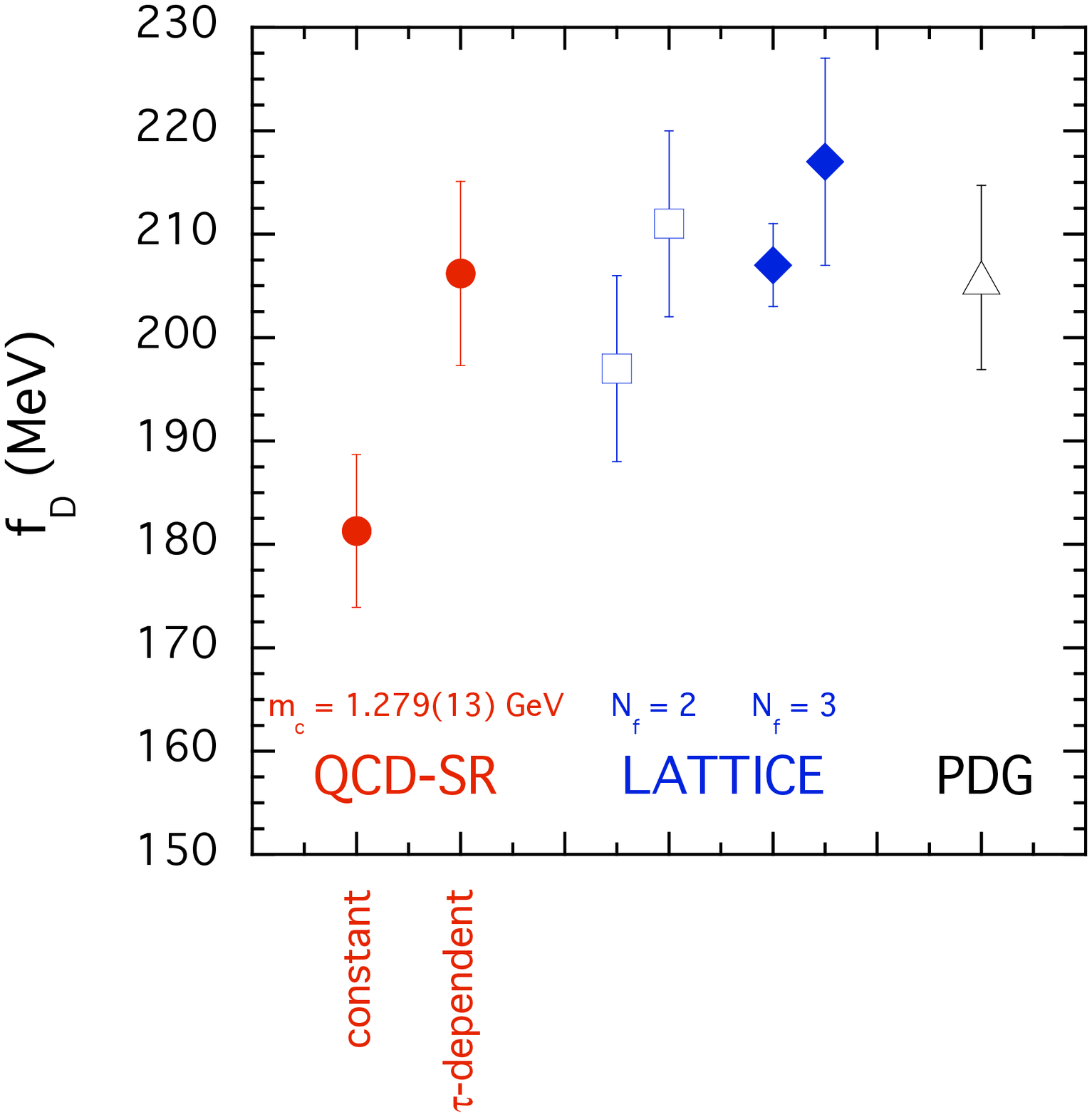}&
\includegraphics[height=7cm]{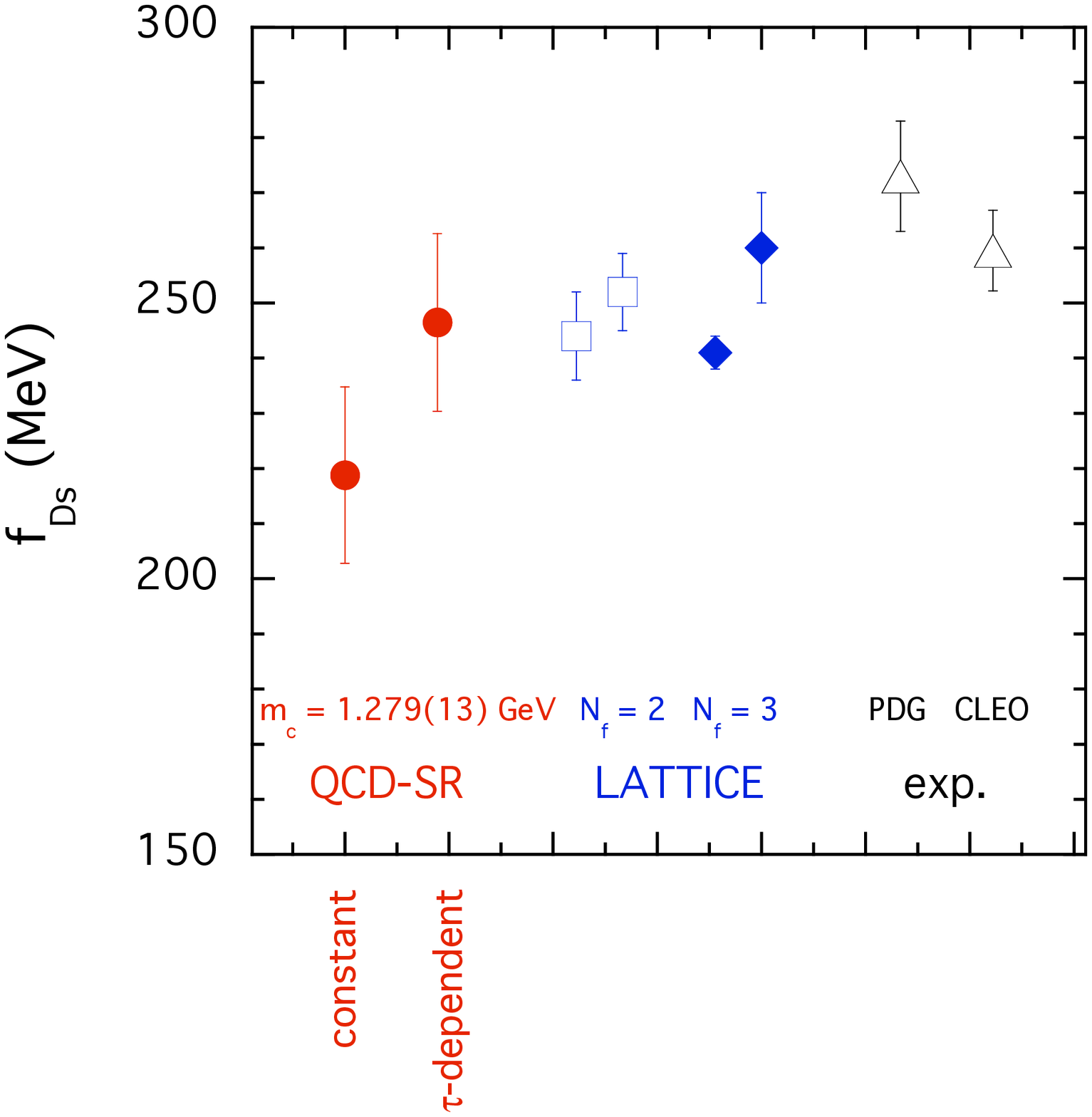}
\end{tabular}
\caption{\label{Plot:Dresults}Comparison of our results for $f_D$
and $f_{D_s}$ with lattice results; for a detailed list of
references,~cf.~\cite{lms2010}.}
\end{center}
\end{figure}

The application of our extraction procedures leads to 
the following results for charmed mesons:
\begin{eqnarray}
\label{Dresults} f_{D} &=& (206.2 \pm 7.3_{\rm (OPE)} \pm 5.1_{\rm
(syst)})\; \mbox{MeV},\\ f_{D_s} &=& (245.3 \pm 15.7_{\rm (OPE)}
\pm 4.5_{\rm (syst)})\; {\rm MeV}.
\end{eqnarray}
The OPE-related error is obtained by the bootstrap allowing for the
variation of all QCD parameters (i.e., quark masses, $\alpha_s$,
condensates) in the relevant ranges. One observes a perfect
agreement~of~our predictions with the respective lattice results
(Fig.~\ref{Plot:Dresults}). It should be emphasized that our
$\tau$-dependent threshold is a crucial ingredient for a
successful extraction of the decay constants from the sum rule.
Obviously, the standard $\tau$-independent approximation yields a
much lower value for $f_D$ lying rather far from the experimental
data {\em and\/} from the lattice results.
\vspace{-.5cm}
\section{Decay constants of the $B$ and $B_s$ mesons}
\begin{figure}[!t]
\begin{center}
\begin{tabular}{cc}
\includegraphics[height=6cm]{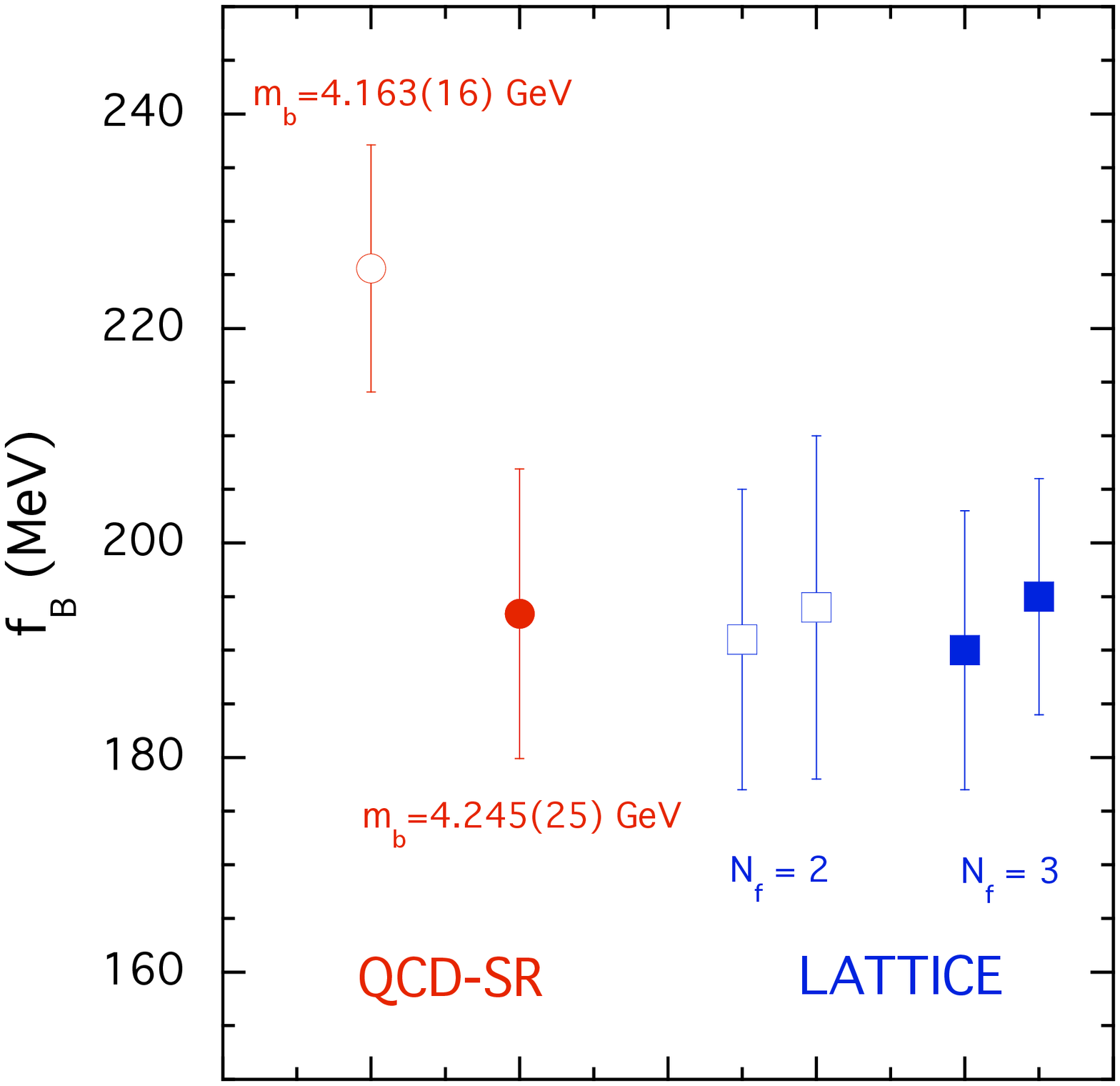}&
\includegraphics[height=6cm]{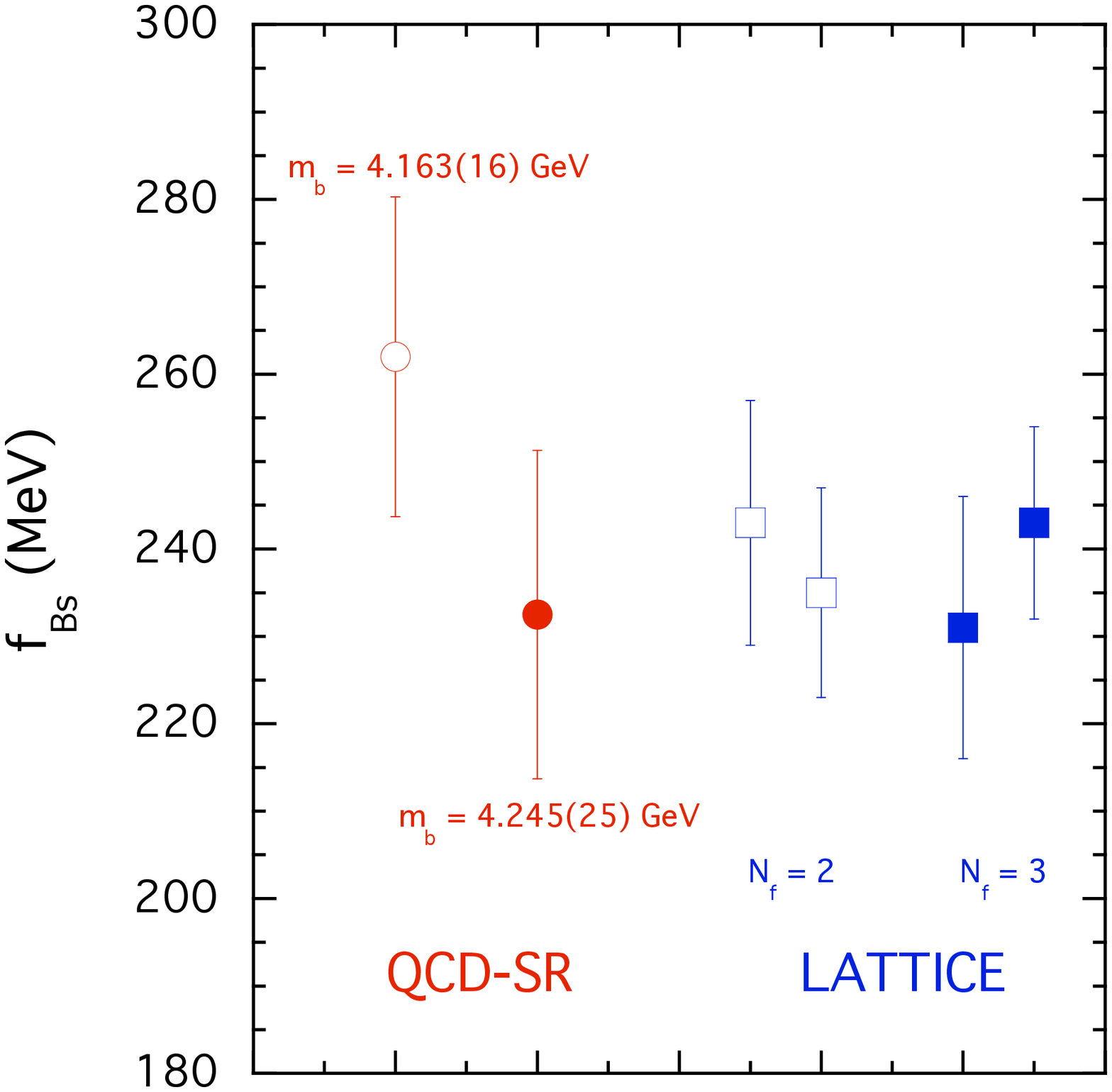}
\end{tabular}
\end{center}
\caption{\label{Plot:Bresults}Comparison of our results for $f_B$
and $f_{B_s}$ with lattice results; for a detailed list of references,~cf.~\cite{lms2010}.}
\end{figure}

The values of the beauty-meson decay constants extracted from QCD
sum rules are extremely sensitive to the precise value of
$\overline{m}_b(\overline{m}_b)$. For instance, the range
$\overline{m}_b(\overline{m}_b)=(4.163\pm
0.016)\;\mbox{GeV}$~\cite{mb} yields results for the decay
constants that are barely compatible with the lattice
calculations~(Fig.~\ref{Plot:Bresults}). Requiring our sum-rule result for $f_B$ 
to match the average of the lattice determinations entails a rather precise value of the $b$-quark
mass
\begin{eqnarray}
\overline{m}_b(\overline{m}_b)=(4.245\pm 0.025)\; {\rm GeV}.
\end{eqnarray}
Our sum-rule estimates for $f_B$ and $f_{B_s}$ corresponding to this value 
of the $b$-quark mass read
\begin{eqnarray}
\label{Bresults} f_{B} &=& (193.4 \pm 12.3_{\rm (OPE)} \pm
4.3_{\rm (syst)})\; {\rm MeV},\\ f_{B_s} &=& (232.5 \pm 18.6_{\rm
(OPE)} \pm 2.4_{\rm (syst)})\; {\rm MeV}.
\end{eqnarray}
\newpage
\vspace{-.3cm}
\section{Conclusions}
\noindent 
1. Our study of {\em charmed mesons\/} clearly demonstrates that the
use of Borel-parameter-dependent thresholds leads to two essential
improvements: (i) The accuracy of decay constants extracted~from
sum rules is considerably improved. (ii) It has become possible to
obtain realistic systematic~errors and to reduce their values to
the level of a few percent. The application of our prescription 
brings~the QCD sum-rule results into perfect agreement with the
findings of both lattice QCD and experiment.

\vspace{.3cm}
\noindent 
2. The {\em beauty-meson\/} decay constants are extremely sensitive to the
precise value of $\overline{m}_b(\overline{m}_b)$; matching the
results from QCD sum rules for $f_B$ to the average of the lattice
evaluations allows us to provide~a rather accurate estimate of the
$b$-quark mass. Our $m_b$ value is in good agreement with several
lattice results but, interestingly, does not overlap with the recent accurate
determination \cite{mb} (for details, consult Ref.~[1]).

\vspace{.5cm}
\noindent
{\bf Acknowledgments.} D.M.\ was supported by the Austrian Science Fund (FWF), project no.~P20573.

\end{document}